\documentclass[aps,prb,reprint,twocolumn,superscriptaddress,showpacs,floatfix,longbibliography]{revtex4-1}

\usepackage{amsmath}
\usepackage{color}
\usepackage{mathrsfs}
\usepackage{dcolumn}
\usepackage{bm}
\usepackage{multirow}
\usepackage{graphicx}
\usepackage{rotating}
\usepackage{layout}
\usepackage[version=3]{mhchem}
\usepackage{braket}
\usepackage{setspace}
\usepackage{comment}

\allowdisplaybreaks
\usepackage{xfrac}
\usepackage{mathtools}
\usepackage{mathrsfs}
\usepackage{subcaption}

\usepackage[colorlinks,citecolor=blue,urlcolor=blue,bookmarks=false,hypertexnames=true]{hyperref}

\begin{document}
\title{Quantum resource reduction for quantum-centric supercomputing via correlated mean-field downfolding framework}
\date{\today}
\author{Thien Ngoc Tran}
\affiliation{Faculty of Physics and Engineering Physics, University of Science, Ho Chi Minh City 700000, Vietnam}
\affiliation{Vietnam National University, Ho Chi Minh City 700000, Vietnam}
\author{Lan Nguyen Tran}
\email{tnlan@hcmus.edu.vn}
\affiliation{Faculty of Physics and Engineering Physics, University of Science, Ho Chi Minh City 700000, Vietnam}
\affiliation{Vietnam National University, Ho Chi Minh City 700000, Vietnam}
\date{\today}

\begin{abstract}
We present OBDF-SQD, a hybrid quantum-classical method that combines one-body downfolding~(OBDF) based on
one-body M\o{}ller--Plesset second-order perturbation
theory~(OBMP2) with sample-based quantum diagonalization~(SQD) for use in quantum-centric supercomputing~(QCS). In this approach, OBMP2 is
executed classically to fold dynamical correlation from external orbitals into a renormalized one-body operator, yielding an effective active-space
Hamiltonian that retains the same operator structure as the bare Hamiltonian and therefore requires no additional quantum circuit resources. SQD is then applied to this effective Hamiltonian, where, 
in this work, the quantum sampling is performed via 
the Qiskit Aer simulator rather than actual quantum hardware. We benchmark OBDF-SQD on dissociation curves of \ce{H6} chain, ring, and lattice systems and the \ce{N2} molecule in the cc-pVDZ basis, comparing against standard methods and active-space SQD (CAS-SQD). We observed that OBDF-SQD consistently improves upon CAS-SQD with the same active space. The
simplicity of the one-body downfolding correction also makes the approach straightforwardly extensible to periodic solids within existing quantum embedding frameworks.
\end{abstract}

\maketitle

\section{Introduction}

Quantum computing is highly promising for simulating complicated molecules and materials challenging to the classical methods ~\cite{bauer2020quantum,mcardle2020quantum,cao2019quantum,motta2021emerging}. 
A central vision guiding current development is that of \emph{quantum-centric supercomputing} (QCS)~\cite{bravyi2022future,kim2023evidence}, in which quantum processors are tightly integrated with classical high-performance computing (HPC) resources. In this framework, quantum hardware serves as a 
specialized co-processor that handles computationally intractable tasks, while 
classical supercomputers manage the remaining workload. This framework is particularly 
well-suited to near-term noisy intermediate-scale quantum (NISQ) devices, which remain limited by noise and decoherence~\cite{de2021materials}, and motivates the development of hybrid quantum-classical algorithms that are robust to hardware 
imperfections and can scale alongside improvements in both quantum and classical 
computing.

A prominent family of hybrid quantum-classical algorithms is built around the 
variational quantum eigensolver 
(VQE)~\cite{peruzzo2014variational,mcclean2016theory,kandala2017hardware}, in which 
a parameterized quantum circuit prepares a trial state and the expectation value of 
the Hamiltonian is measured on the quantum device, while a classical optimizer updates 
the circuit parameters. The unitary coupled-cluster (UCC) 
ansatz~\cite{bartlett1989alternative,taube2006new,romero2018strategies,anand2022quantum} 
has been widely adopted as a chemically motivated state-preparation ansatz for VQE. 
Substantial effort has been devoted to reducing the quantum resources required by VQE, such as qubit coupled-cluster (QCC)\cite{QCC2018,QCC2020}, ADAPT-VQE~\cite{grimsley2019adaptive,tang2021qubit,zhang2021mutual}, and factorized UCC ansatz~\cite{fac-UCC,xu2022decomposition}. Other 
adaptive schemes inspired by classical quantum chemistry have also been 
proposed~\cite{selectiveUCC2022,projectiveVQE2021}. Complementary strategies reduce 
the problem size by partitioning the system into active spaces or subsystems. 
Nakagawa and 
co-workers proposed deep VQE~\cite{fujii2022deep,mizuta2021deep}, dividing the system 
into subsystems each solved independently with VQE. Quantum embedding frameworks, 
including density matrix embedding theory 
(DMET)~\cite{mineh2022solving,tilly2021reduced,Knizia2013}, dynamical 
mean-field theory (DMFT)~\cite{keen2020quantum,bauer2016hybrid}, and density 
functional embedding~\cite{Dresselhaus2015,Gujarati2023}, have further 
extended the reach of VQE to larger systems. Despite these advances, VQE faces 
persistent practical difficulties: the classical optimization loop is costly, 
convergence can be hindered by barren plateaus, and circuit depths required for 
chemically accurate results remain challenging on current hardware.

Sample-based quantum diagonalization  (SQD)~\cite{RobledoMoreno2025,Barison2025,Danilov2025,Kaliakin2025,Shajan2025} offers an alternative route to hybrid quantum-classical computation that resolves many of these difficulties. Rather than optimizing a parameterized 
circuit, SQD leverages repeated projective measurements, bitstring samples, obtained 
from a quantum device to construct a compressed configuration-interaction (CI) 
subspace, which is then diagonalized classically to approximate the ground-state 
energy. Because SQD requires no classical optimization of circuit parameters and 
naturally delegates the diagonalization step to classical HPC resources, it aligns 
well with the QCS model. SQD has been demonstrated on real quantum hardware to 
achieve chemically accurate results for molecules such as the nitrogen dimer and 
iron-sulfur clusters~\cite{RobledoMoreno2025}, systems that pose a formidable 
challenge to classical methods.

A key limitation shared by both VQE and SQD in active-space formulations is the 
incomplete treatment of dynamical electron correlation outside the active space. 
Neglecting this contribution leads to systematic errors in energies, particularly 
for small active spaces. To address this challenge, Kowalski and co-workers employed downfolding based on the double UCC (DUCC) ansatz to construct effective active-space Hamiltonians~\cite{bauman2019downfolding,kowalski2021dimensionality} that integrate 
out high-energy degrees of freedom while reproducing exact energies. In the same vein, we have recently developed the one-body downfolding (OBDF) in a resource-efficient way\cite{OBMP2-JPCA2023}. In the OBDF framework based on one-body M{\o}ller-Plesset second-order perturbation theory 
(OBMP2)~\cite{OBMP2-JCP2013,OBMP2-JPCA2021,OBMP2-PCCP2022,OBMP2-JPCA2024,OBMP2-JCP2025}, the effect of external (inactive) orbitals is incorporated into the active-space Hamiltonian through a 
correlated one-electron operator that modifies only the one-body part of the 
Hamiltonian. The resulting effective Hamiltonian retains the same operator 
structure as the bare active-space Hamiltonian and therefore requires no 
additional quantum resources beyond those needed for the bare problem. OBDF can be derived from the DUCC ansatz by restricting the external cluster operator to double excitations involving at least one inactive index and truncating the Baker-Campbell-Hausdorff (BCH) expansion at second 
order. This makes OBDF a natural and practical component within the QCS framework: the additional computational cost is entirely classical, while the quantum circuit complexity is unchanged.

In this work, we combine SQD with the OBDF framework (OBDF-SQD) to simultaneously address noise tolerance, active-space accuracy, and dynamical correlation. The method proceeds as follows: OBMP2 is first run classically for the full 
system to generate correlated molecular orbitals and to construct the effective active-space Hamiltonian encoding dynamical correlation outside the active space through a modified one-body term; SQD is then applied to this effective Hamiltonian, with the quantum 
sampling performed through the Qiskit Aer simulator in place of real quantum hardware. We 
benchmark OBDF-SQD on the dissociation of H$_6$ chain, ring, and lattice in the cc-pVDZ basis set, as well as the N$_2$ molecule, comparing against FCI, CCSD(T), HF, OBMP2, and standard CAS-SQD. Our results demonstrate that OBDF-SQD systematically improves upon CAS-SQD with the same active space. We also discuss the current limitations of the OBDF framework 
and highlight promising directions for future work.

\section{Theory}

\subsection{Sample-Based Quantum Diagonalization}
\label{sec:sqd}

The many-body electronic Hamiltonian in second quantization reads
\begin{equation}
    \hat{H} = \sum_{pq\sigma} h_{pq}\, \hat{a}^{\dagger}_{p\sigma}\hat{a}_{q\sigma} 
    + \frac{1}{2}\sum_{pqrs\,\sigma\eta}g_{pq}^{rs}\, 
    \hat{a}^{\dagger}_{p\sigma}\hat{a}^{\dagger}_{r\eta}\hat{a}_{q\sigma}\hat{a}_{s\eta},
    \label{eq:hamiltonian}
\end{equation}
where $h_{pq}$ and $g_{pq}^{rs}$ are the one- and two-electron integrals, 
respectively. The Jordan--Wigner (JW) transformation maps the fermionic 
operators onto Pauli strings acting on $2N_\mathrm{MO}$ qubits, with 
computational basis states corresponding to Slater determinants encoded 
as bitstrings $|x\rangle = (x_{\downarrow}, x_{\uparrow})$, where 
$x_{p\sigma} \in \{0,1\}$ denotes the occupation of the $p$-th spin orbital.

In the SQD framework \cite{RobledoMoreno2025,Barison2025,Danilov2025,Kaliakin2025,Shajan2025}, quantum circuits 
employing the Local Unitary Cluster Jastrow (LUCJ) ansatz \cite{Motta2023} 
are used to generate a set of sampled bitstring configurations 
$\tilde{\mathcal{X}}$. In this work, the quantum sampling is performed 
via the Qiskit Aer simulator as an emulator of physical quantum hardware\cite{Qiskit}. 
Particle-number conservation, $N_x = N$, is enforced to filter the valid 
subset $\mathcal{X}_N \subset \tilde{\mathcal{X}}$. Configurations 
violating this constraint are subjected to a self-consistent recovery 
procedure, wherein bits are probabilistically flipped according to a 
modified ReLU weighting function\cite{RobledoMoreno2025}.

The recovered configurations are partitioned into $K$ batches, each 
defining a subspace $\mathcal{S}^{(k)}$ spanned by $d$ bitstrings. 
The Hamiltonian is projected onto each subspace as
\begin{equation}
    \hat{H}_{\mathcal{S}^{(k)}} = \hat{P}_{\mathcal{S}^{(k)}} 
    \hat{H} \hat{P}_{\mathcal{S}^{(k)}}, \quad 
    \hat{P}_{\mathcal{S}^{(k)}} = \sum_{x \in \mathcal{S}^{(k)}} |x\rangle\langle x|,
    \label{eq:projection}
\end{equation}
and diagonalized to yield the subspace ground-state energy $E^{(k)}$ 
and wave function $|\psi^{(k)}\rangle$. The average spin-orbital 
occupation is updated as
\begin{equation}
    n_{p\sigma} = \frac{1}{K}\sum_{k=1}^{K} 
    \langle \psi^{(k)}|\hat{n}_{p\sigma}|\psi^{(k)}\rangle,
    \label{eq:avg_occ}
\end{equation}
and the procedure iterates until $n_{p\sigma}$ converges.

\subsection{OBDF-SQD framework}

The OBMP2 approach~\cite{OBMP2-JCP2013,OBMP2-JPCA2021} is derived through the 
canonical transformation~\cite{CT-JCP2006,CT-JCP2007,CT-ACP2007,CT-JCP2009,
CT-JCP2010,CT-IRPC2010}, in which an effective Hamiltonian incorporating dynamic 
correlation is obtained via a similarity transformation of the molecular Hamiltonian 
$\hat{H}$ using a unitary operator $e^{\hat{A}}$:
\begin{align}
\hat{\bar{H}} = e^{\hat{A}^\dagger} \hat{H} e^{\hat{A}},
\label{Hamiltonian:ct}
\end{align}
where the anti-Hermitian operator $\hat{A} = \hat{A}_\mathrm{D}$ includes only double 
excitations with MP2 amplitudes
\begin{align}
  T_{i j}^{a b} =  \frac{g_{i j}^{a b}}{\epsilon_{i} + \epsilon_{j} - \epsilon_{a} - \epsilon_{b}}.
  \label{eq:amp}
\end{align}
Using the Baker--Campbell--Hausdorff (BCH) expansion and the cumulant 
approximation~\cite{cumulant-JCP1997,cumulant-PRA1998,cumulant-CPL1998,
cumulant-JCP1999} to retain only one-body contributions, the resulting OBMP2 
Hamiltonian takes the form
\begin{align}
  \hat{H}_\text{OBMP2} = \hat{H}_\text{HF} + \hat{v}_\text{OBMP2},
  \label{eq:h4}
\end{align}
where $\hat{H}_\text{HF} = \hat{F} + C$ is the Hartree--Fock Hamiltonian and 
$\hat{v}_\text{OBMP2}$ is a correlated one-body potential whose explicit expression 
is given in Refs.~\citenum{OBMP2-JCP2013,OBMP2-JPCA2021}. Equation~\eqref{eq:h4} 
can be written compactly as
\begin{align}
  \hat{H}_\text{OBMP2} = \hat{\bar{F}} + \bar{C},
  \label{eq:h5}
\end{align}
with the correlated Fock matrix
\begin{align}
\bar{f}^{p}_{q} = f^{p}_{q} + v^{p}_{q},
\label{eq:corr-fock}
\end{align}
where $v^{p}_{q}$ is the matrix representation of $\hat{v}_\text{OBMP2}$. 
Diagonalizing $\bar{f}^{p}_{q}$ yields a set of correlated molecular orbitals 
that implicitly encode the effect of dynamic correlation on the one-body picture. 
The formal scaling of OBMP2 is $\mathcal{O}(N^5)$, analogous to standard MP2.

To construct an effective active-space Hamiltonian, we adopt the double unitary 
coupled-cluster (DUCC) 
framework~\cite{bauman2019downfolding,kowalski2021dimensionality}. Partitioning 
the full orbital space into active and inactive subspaces, the DUCC ansatz is written as
\begin{equation}
    |\phi\rangle = e^{\hat{A}_\mathrm{ext}}\, e^{\hat{A}_\mathrm{int}}\, |0\rangle,
    \label{eq:ducc}
\end{equation}
where $\hat{A}_\mathrm{int}$ generates excitations within the active space and 
$\hat{A}_\mathrm{ext}$ generates excitations involving at least one inactive orbital. 
The total energy can then be expressed as
\begin{align}
    E = \langle 0 | e^{\hat{A}^\dagger_\mathrm{int}}\, \bar{\hat{H}}_\mathrm{ext}\, 
    e^{\hat{A}_\mathrm{int}} | 0 \rangle,
    \label{eq:total_energy}
\end{align}
where the downfolded Hamiltonian $\bar{\hat{H}}_\mathrm{ext}$ is defined by
\begin{equation}
    \bar{\hat{H}}_\mathrm{ext} = e^{\hat{A}^\dagger_\mathrm{ext}} \hat{H}\, 
    e^{\hat{A}_\mathrm{ext}} \simeq \hat{H} + \left[\hat{H}, \hat{A}_\mathrm{ext}\right] 
    + \frac{1}{2}\left[\left[\hat{H}, \hat{A}_\mathrm{ext}\right], \hat{A}_\mathrm{ext}\right],
    \label{eq:eff_ham}
\end{equation}
with the BCH expansion truncated at second order. Restricting 
$\hat{A}_\mathrm{ext} = \hat{A}^D_\mathrm{ext}$ to double excitations carrying at 
least one inactive orbital index and invoking the OBMP2 approximation for the 
commutator terms in Eq.~\eqref{eq:eff_ham}, the downfolded Hamiltonian becomes
\begin{equation}
    \bar{\hat{H}}_\mathrm{ext} = \hat{H} + \hat{v}^\mathrm{ext}_\mathrm{OBMP2},
    \label{eq:downfolded_ham}
\end{equation}
where $\hat{v}^\mathrm{ext}_\mathrm{OBMP2}$ is the external correlation potential 
constructed from amplitudes carrying at least one inactive orbital index. Projecting 
onto the active space, Eq.~\eqref{eq:downfolded_ham} yields the one-body downfolded 
(OBDF) active-space Hamiltonian
\begin{equation}
    \hat{H}_\mathrm{OBDF} = \hat{H}_\mathrm{CAS} + \hat{v}^\mathrm{ext}_\mathrm{OBMP2},
    \label{eq:obdf_ham}
\end{equation}
where $\hat{H}_\mathrm{CAS}$ is the bare active-space Hamiltonian. Since 
$\hat{v}^\mathrm{ext}_\mathrm{OBMP2}$ is a one-body operator, $\hat{H}_\mathrm{OBDF}$ 
retains the same two-body operator structure as $\hat{H}_\mathrm{CAS}$, requiring 
no additional quantum resources beyond those needed for the bare active-space problem.

Within this framework, SQD is employed to solve the eigenvalue problem defined by 
$\hat{H}_\mathrm{OBDF}$. The OBMP2 procedure first provides a set of correlated 
molecular orbitals, obtained by diagonalizing the correlated Fock matrix 
$\bar{f}^{q}_{p}$ in Eq.~\eqref{eq:corr-fock}, which serve as the one-body basis 
for the active space. The OBDF active-space Hamiltonian in Eq.~\eqref{eq:obdf_ham} 
is then constructed in this correlated orbital basis. SQD is subsequently applied 
to $\hat{H}_\mathrm{OBDF}$ following the procedure described in 
Section~\ref{sec:sqd}: bitstring samples from quantum circuit measurements are 
used to define a subspace of Slater determinants within the active space, and the 
Hamiltonian is constructed and diagonalized within this subspace classically. 
Using correlated orbitals as the single-particle reference ensures that the 
sampled configurations are more representative of the true correlated ground state, 
improving the quality of the SQD subspace and the accuracy of the resulting energies 
and one-particle density matrices. This combination of OBMP2 downfolding and SQD, 
hereafter referred to as OBDF-SQD, constitutes the central methodological 
contribution of the present work. 

OBMP2 and OBDF are implemented within a local version of PySCF~\cite{pyscf-2018}. SQD is executed using the 
Qiskit-based quantum circuit simulator \cite{Qiskit} deployed on a High-Performance 
Computing (HPC) system. All standard methods have been performed using pySCF.

\section{Numerical demonstration}
\label{sec:results}
\subsection{H$_6$ systems}
\begin{figure}[t]
    \centering
\includegraphics[width=1.0\columnwidth, height = 0.4\textheight]{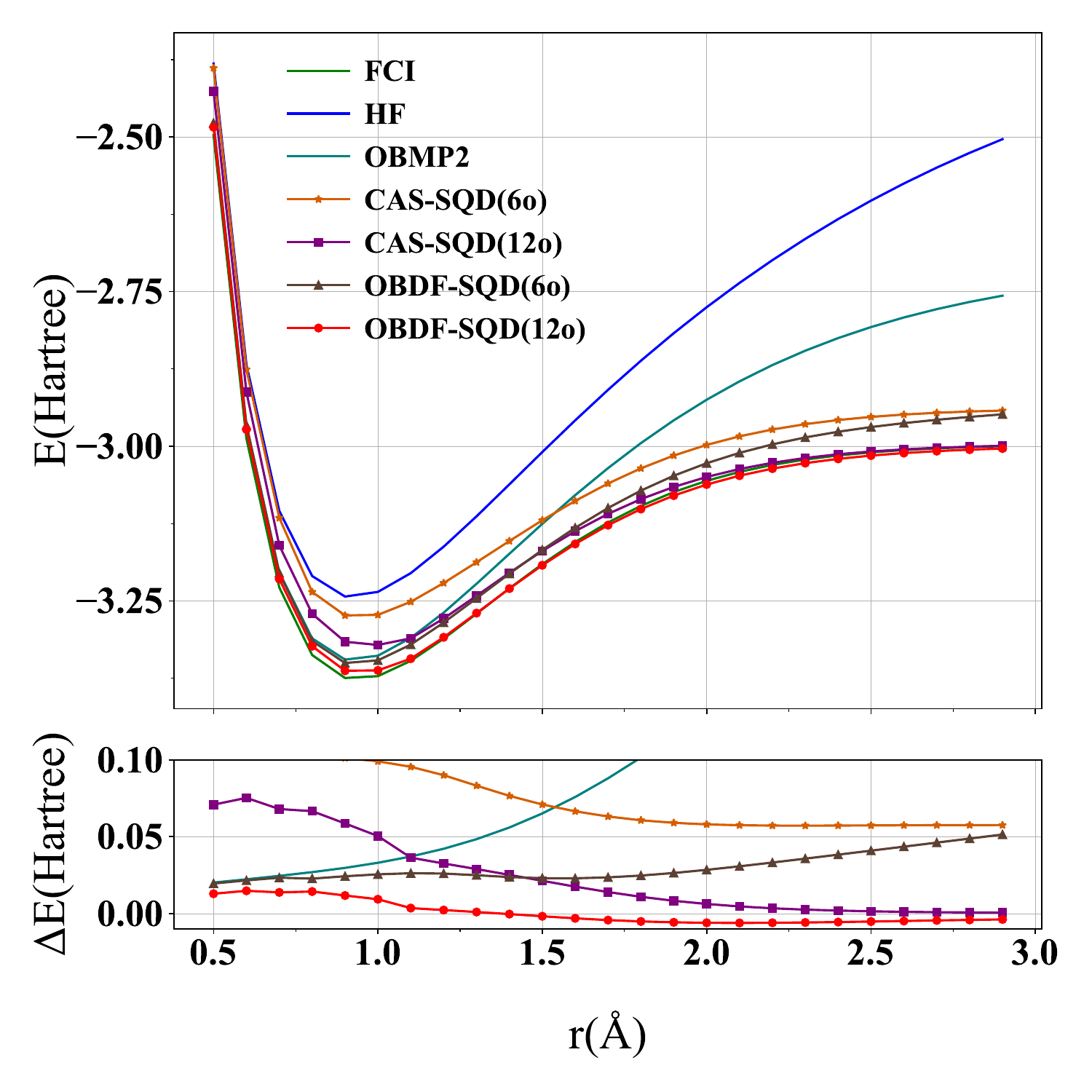}
    \caption{Upper: potential energy curves of H$_6$ chain in cc-pVDZ. Lower: energy errors relative to FCI.}
    \label{fig:h6chain}
\end{figure}

\begin{figure}[t]
    \centering
    \hspace{-5mm}
    \includegraphics[width=1.0\columnwidth, height = 0.4\textheight]{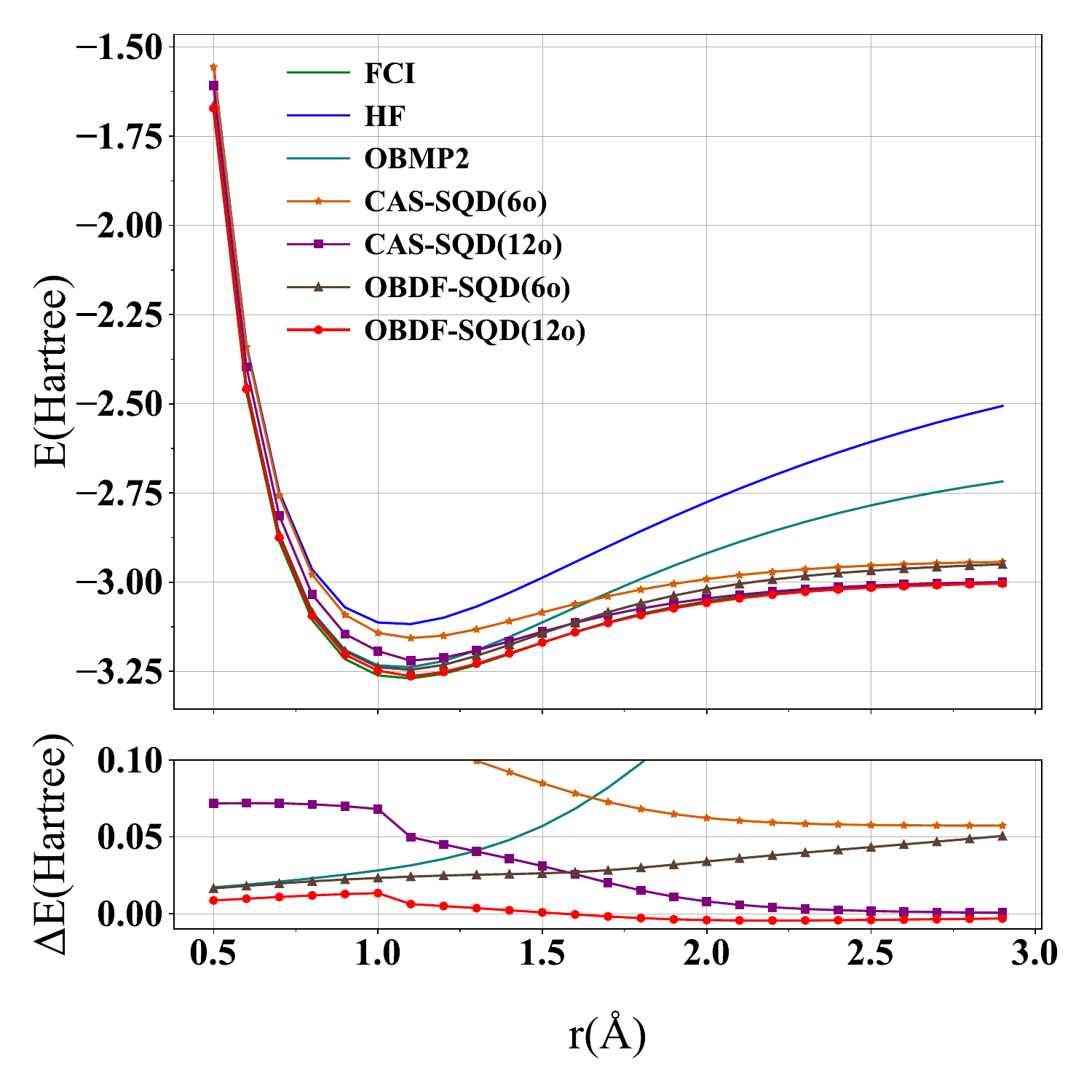}
    \caption{Upper: potential energy curves of H$_6$ lattice in cc-pVDZ. Lower: energy errors relative to FCI.}
    \label{fig:h6lattice}
\end{figure}

\begin{figure}[t]
    \centering
    \hspace{-5mm}
    \includegraphics[width=1.0\columnwidth, height = 0.4\textheight]{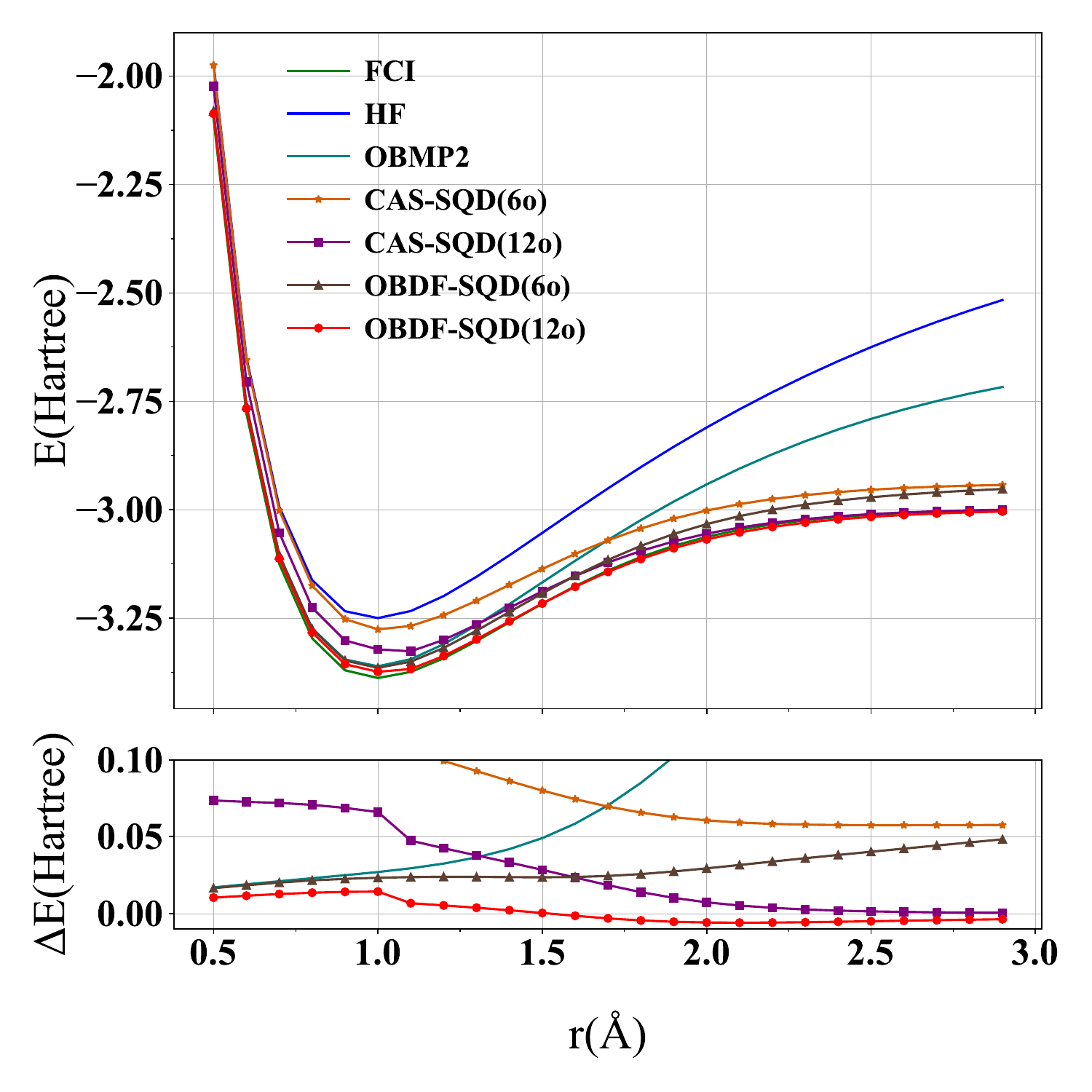}
    \caption{Upper: potential energy curves of H$_6$ ring in cc-pVDZ. Lower: energy errors relative to FCI.}
    \label{fig:h6ring}
\end{figure}

To benchmark the performance of the OBDF-SQD method, we computed 
potential energy curves (PECs) for three H$_6$ geometries, linear chain, lattice, and ring. OBDF-SQD results are compared against Hartree-Fock (HF), OBMP2, conventional SQD within complete active spaces (CAS-SQD) with active spaces of 6 orbtials (6o) and 12 orbitals (12o), and full configuration interaction (FCI) as the exact reference. The absolute energies are presented in the upper panels of Figs.~\ref{fig:h6chain}--\ref{fig:h6ring}, and the corresponding errors with respect to FCI are displayed in the lower panels.

We first evaluate the PEC of the H$_6$ linear chain (Fig.~\ref{fig:h6chain}). At short bond lengths, the electronic structure is dominated by dynamical correlation, and the wavefunction retains a predominantly single-reference character. In this regime, all methods yield energies in reasonable agreement with FCI. However, a closer look at the error panel reveals important distinctions among the methods. Importantly, OBDF-SQD with both active spaces of 6o and 12o demonstrates a notably smaller energy error relative to FCI compared to CAS-SQD. This enhanced accuracy in the short-distance regime is thanks to the one-body downfolding transformation, which systematically incorporates contributions of virtual orbitals outside the active space into an effective renormalized Hamiltonian. 
At compressed geometries, the one-body correlated potential underlying the downfolding is sufficient, allowing the effective Hamiltonian to accurately capture dynamical correlation effects that are excluded from the active space in a conventional CAS-SQD treatment. 

As the interatomic distance is increased beyond $r \approx 1.5$~\AA\ and the system enters the strongly correlated regime, marked differences emerge among the methods. The HF method fails dramatically at large bond lengths, exhibiting a well-known unphysical increase in energy due to its inability to properly describe bond dissociation and the associated multireference character. OBMP2 improves upon HF in the weakly correlated region but also deteriorates at stretched geometries, where the one-body and perturbative treatments are insufficient to capture the strong correlation. Both CAS-SQD(6o) and OBDF-SQD(6o) yield considerably improved PECs relative to HF and OBMP2, but the errors are still quite large. When the distance increases beyond $2.5$~\AA\, OBDF-SQD(6o) demonstrates accuracy marginally superior to CAS-SQD(6o), implying that the correlation captured by OBDF is insufficient to reduce the error. Upon expanding the active orbital space to 12o, a substantial improvement is observed for both methods. CAS-SQD(12o) and OBDF-SQD(12o) both are very close to the FCI reference with small errors.

The PECs of the H$_6$ lattice (Fig.~\ref{fig:h6lattice}) and H$_6$ ring (Fig.~\ref{fig:h6ring}) geometries show trends consistent with those observed for the linear chain. At short interatomic bond lengths, OBDF-SQD consistently outperforms CAS-SQD for both active space sizes, yielding smaller energy errors relative to FCI. As in the case of H chain, the one-body downfolding transformation incorporates dynamical correlation contributions from external orbitals into an effective renormalized Hamiltonian, which is particularly beneficial in the single-reference regime where the perturbative treatment is well-conditioned. 
The advantage of OBDF-SQD over CAS-SQD at short bond lengths is significantly notable even with the 6-orbital active space, indicating that the downfolding renormalization captures a non-negligible fraction of dynamical correlation that would otherwise be absent from a conventional active space treatment.

As the interatomic distance is stretched into the strongly correlated regime, HF and OBMP2 again fail to provide qualitatively correct descriptions of dissociation, consistent with the chain results. For the 6-orbital active space, both OBDF-SQD(6o) and CAS-SQD(6o) show substantial deviations from FCI at large bond lengths. Although OBDF-SQD(6o) demonstrates lower errors than CAS-SQD(6o) across all bond lengths considered here, at very large interatomic distances, the two methods appear to converge toward comparable error magnitudes. This is because, while the one-body correlated potential is effective at capturing dynamical correlation, it remains insufficient on its own to fully describe the strong correlation that dominates beyond the dissociation limit when the active space is restricted to 6 orbitals. As observed for the chain, upon expanding the active space to 12 orbitals, both CAS-SQD(12o) and OBDF-SQD(12o) achieve near-FCI accuracy at stretched bond lengths.
 
\subsection{N$_2$ molecule}

\begin{figure}[t]
    \centering
    \hspace{-5mm}
    \includegraphics[width=1.0\columnwidth, height = 0.4\textheight]{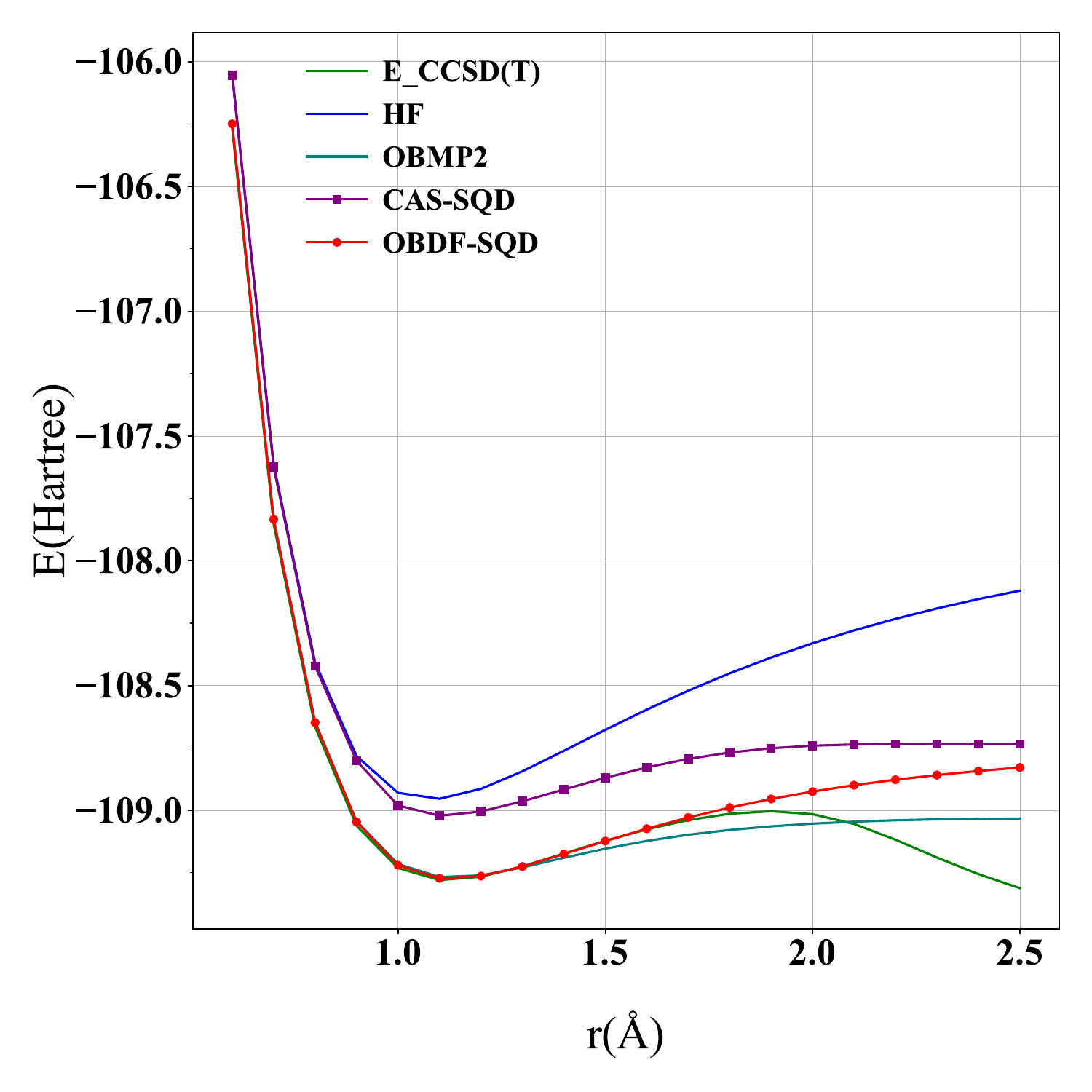}
    \caption{Potential energy curves of N$_2$ ring in cc-pVDZ.}
    \label{fig:n2}
\end{figure}

To assess the performance of OBDF-SQD beyond H systems, we applied the method to the dissociation of N$_2$, a challenging benchmark system in quantum chemistry. Active spaces of 6 orbitals from N 2$p$ are considered for SQD, and the 
results are presented in Fig.~\ref{fig:n2}. CCSD(T), considered as the ``gold standard'' of single-reference 
quantum chemistry, serves as the reference for assessing accuracy in the 
short-distance regime, where the N$_2$ wavefunction is predominantly 
single-reference in character and CCSD(T) is expected to be highly 
accurate. However, as the N--N bond is elongated beyond 
$r \approx 1.5$~\AA, the wavefunction acquires strong multireference 
character and the single-reference coupled-cluster expansion breaks down 
abruptly, causing CCSD(T) to diverge and yield unphysical energies in 
the dissociation region. Consequently, CCSD(T) cannot serve as a 
reliable reference at large bond lengths.

As seen in Figure~\ref{fig:n2}, HF fails to provide a qualitatively correct description across the entire dissociation curve, 
yielding energies that lie well above those of all correlated methods 
at all geometries. OBMP2 recovers a substantial portion of the dynamical correlation energy 
and agrees reasonably well with CCSD(T) near the equilibrium geometry. As we discuss previously\cite{OBMP2-JPCA2024}, OBMP2 does not diverge abruptly in the stretched region. This smoother behavior is due to the self-consistency of the orbital optimization in OBMP2, which provides a degree of robustness against the near-degeneracy of orbital energies. Nevertheless, OBMP2 predicts energies unphysically low at large bond lengths, indicating that it remains unreliable for a quantitative description of bond dissociation.

Both CAS-SQD and OBDF-SQD remain well-behaved and non-divergent across the entire range of bond lengths studied, properly capturing the dissociation. 
In the short-distance regime where CCSD(T) is reliable, OBDF-SQD consistently yields energies closer to CCSD(T) than CAS-SQD. Also, it provides a more physically correct dissociation curve at large bond lengths where CCSD(T) fails. The improvement of OBDF-SQD over CAS-SQD is most pronounced at short distances, where dynamical correlation is dominant. 
At large bond lengths, however, the gap between OBDF-SQD(6o) and 
CAS-SQD(6o) progressively narrows, and the two methods seem to converge together in the strongly correlated dissociation regime. As the N--N bond is stretched, the electronic structure becomes increasingly dominated by strong static correlation arising from the near-degeneracy of multiple electronic configurations. Consequently, the effective 
one-body renormalization provided by the downfolding procedure 
is insufficient to capture this strong correlation, and OBDF-SQD 
gradually reduces to CAS-SQD in this limit. 
Incorporating higher-order downfolding or adaptive active-space selection would be needed to reach higher accuracies in the strongly correlated regions.

\section{Conclusion and outlook}
We have introduced and benchmarked OBDF-SQD, a hybrid quantum-classical framework that integrates one-body downfolding with sample-based quantum
diagonalization. The method is motivated by the QCS approach, in
which quantum processors and classical HPC resources are used in concert, each handling the tasks for which it is
best suited. In OBDF-SQD, the OBMP2 pre-processing and the final CI diagonalization are performed entirely on classical hardware, while the quantum device, simulated on HPC in this work using Qiskit emulator, is used solely for bitstring sampling on an effective Hamiltonian that carries no additional circuit complexity relative to bare SQD. This
strict separation reduces the quantum resource requirement without sacrificing the physical information content of each quantum circuit
execution.

Benchmark calculations on \ce{H6} chain, ring, and lattice systems and the \ce{N2} molecule demonstrate that OBDF-SQD consistently outperforms CAS-SQD with the same active space since the renormalized one-body operator efficiently captures active-external dynamical correlation at no additional quantum cost. Expanding the active space to 12~orbitals further reduces errors across the full dissociation curve of H systems.

Several limitations of the present approach, however, should be noted. First, the one-body downfolding correction relies on a perturbative
OBMP2 treatment of the active-external correlation, which becomes unreliable when the reference wavefunction is strongly multireference in character. As a result, in the strongly correlated dissociation
regime, the improvement of OBDF-SQD over CAS-SQD
decreases progressively, and seems that the two methods converge to comparable errors at large bond lengths. Second, the correction modifies only the one-body part of the Hamiltonian and therefore
cannot account for higher-order two-body and many-body active-external correlation effects that become important when the active space is small and the correlation is strong. In such cases, an explicit
enlargement of the active space remains necessary. Third, the current benchmarks are limited to small molecular systems; validation on larger and more diverse systems is needed to establish the
transferability of the method.

Future work could address the identified limitations by incorporating higher-order downfolding corrections, such as effective two-body interactions derived from the double unitary coupled-cluster framework, or by combining OBDF with dynamic active-space selection strategies. Beyond molecular systems, the simplicity of the one-body downfolding framework suggests a natural extension to periodic solids. Because the OBDF
correction requires only a modified one-body operator, it can in principle be applied to solids by extending the underlying OBMP2
formalism to a periodic setting, where $k$-point-resolved orbital energies and two-electron integrals replace their molecular
counterparts. Such an extension would enable OBDF-SQD to treat strongly correlated materials within the QCS framework, combining a
classically constructed periodic OBDF with SQD performed on an effective
impurity or unit-cell Hamiltonian. Working along this direction is currently in progress.

\section*{Acknowledgments}
This research is funded by Vietnam National University,  Ho Chi Minh City (VNU-HCM) under grant number B2026-18-18. The authors used AI-assisted tools only for manuscript language 
editing.

\bibliography{main}

@article{OBMP2-JCP2013,
  title={Correlated one-body potential from second-order M{\o}ller-Plesset perturbation theory: Alternative to orbital-optimized MP2 method},
  author={Tran, Lan Nguyen and Yanai, Takeshi},
  journal={J. Chem. Phys.},
  volume={138},
  number={22},
  pages={224108},
  year={2013},
  publisher={American Institute of Physics},
}

@article{OBMP2-JPCA2021,
  title={Improving perturbation theory for open-shell molecules via self-consistency},
  author={Tran, Lan Nguyen},
  journal={J. Phys. Chem. A},
  volume={125},
  number={41},
  pages={9242},
  year={2021},
  publisher={ACS Publications},
}

@article{OBMP2-PCCP2022,
  title={Can second-order perturbation theory accurately predict electron density of open-shell molecules? The importance of self-consistency},
  author={Tran, Lan Nguyen},
  journal={Phys. Chem. Chem. Phys.},
  volume={24},
  number={32},
  pages={19393},
  year={2022},
  publisher={Royal Society of Chemistry},
}

@article{OBMP2-JPCA2023,
  title={Correlated reference-assisted variational quantum eigensolver},
  author={Le, Nhan Trong and Tran, Lan Nguyen},
  journal={J. Phys. Chem. A},
  volume={127},
  number={24},
  pages={5222},
  year={2023},
  publisher={ACS Publications},
}

@article{OBMP2-JPCA2024,
  title={Reaching high accuracy for energetic properties at second-order perturbation cost by merging self-consistency and spin-opposite scaling},
  author={Tran, Nhan Tri and Nguyen, Hoang Thanh and Tran, Lan Nguyen},
  journal={J. Phys. Chem. A},
  volume={128},
  number={8},
  pages={1543},
  year={2024},
  publisher={ACS Publications},
}

@article{OBMP2-JCP2025,
  title={Attaining high accuracy for charge-transfer excitations in non-covalent complexes at second-order perturbation cost: The importance of state-specific self-consistency},
  author={Tran, Nhan Tri and Tran, Lan Nguyen},
  journal={J. Chem. Phys.},
  volume={162},
  number={10},
  year={2025},
  publisher={AIP Publishing},
}

@article{CT-JCP2006,
  title={Canonical transformation theory for multireference problems},
  author={Yanai, Takeshi and Chan, Garnet Kin-Lic},
  journal={J. Chem. Phys.},
  volume={124},
  number={19},
  pages={194106},
  year={2006},
  publisher={American Institute of Physics}
}

@article{CT-JCP2007,
  title={Canonical transformation theory from extended normal ordering},
  author={Yanai, Takeshi and Chan, Garnet Kin-Lic},
  journal={J. Chem. Phys.},
  volume={127},
  number={10},
  pages={104107},
  year={2007},
  publisher={American Institute of Physics}
}

@incollection{CT-ACP2007,
address = {Hoboken, NJ, USA},
author = {Chan, Garnet Kin-Lic and Yanai, Takeshi},
booktitle = {Adv. Chem. Phys., Volume 134},
chapter = {13},
edition = {Reduced-De},
editor = {Mazziotti, D. A.},
pages = {343},
publisher = {John Wiley \& Sons, Inc.},
title = {{Canonical Transformation Theory for Dynamic Correlations in Multireference Problems}},
year = {2007}
}

@article{CT-JCP2009,
  title={Quadratic canonical transformation theory and higher order density matrices},
  author={Neuscamman, Eric and Yanai, Takeshi and Chan, Garnet Kin-Lic},
  journal={J. Chem. Phys.},
  volume={130},
  number={12},
  pages={124102},
  year={2009},
  publisher={American Institute of Physics}
}

@article{CT-JCP2010,
  title={Strongly contracted canonical transformation theory},
  author={Neuscamman, Eric and Yanai, Takeshi and Chan, Garnet Kin-Lic},
  journal={J. Chem. Phys.},
  volume={132},
  number={2},
  pages={024106},
  year={2010},
  publisher={American Institute of Physics}
}

@article{CT-IRPC2010,
  title={A review of canonical transformation theory},
  author={Neuscamman, Eric and Yanai, Takeshi and Chan, Garnet Kin-Lic},
  journal={Int. Rev. Phys. Chem.},
  volume={29},
  number={2},
  pages={231},
  year={2010},
  publisher={Taylor \& Francis}
}

@article{cumulant-JCP1997,
  title={Normal order and extended Wick theorem for a multiconfiguration reference wave function},
  author={Kutzelnigg, Werner and Mukherjee, Debashis},
  journal={J. Chem. Phys.},
  volume={107},
  number={2},
  pages={432},
  year={1997},
  publisher={American Institute of Physics}
}

@article{cumulant-PRA1998,
  title={Contracted Schr{\"o}dinger equation: Determining quantum energies and two-particle density matrices without wave functions},
  author={Mazziotti, David A},
  journal={Phys. Rev. A},
  volume={57},
  number={6},
  pages={4219},
  year={1998},
  publisher={APS}
}

@article{cumulant-CPL1998,
  title={Approximate solution for electron correlation through the use of Schwinger probes},
  author={Mazziotti, David A},
  journal={Chem. Phys. Lett.},
  volume={289},
  number={5-6},
  pages={419},
  year={1998},
  publisher={Elsevier}
}

@article{cumulant-JCP1999,
  title={Cumulant expansion of the reduced density matrices},
  author={Kutzelnigg, Werner and Mukherjee, Debashis},
  journal={J. Chem. Phys.},
  volume={110},
  number={6},
  pages={2800},
  year={1999},
  publisher={American Institute of Physics}
}

@article{pyscf-2018,
  title={PySCF: the Python-based simulations of chemistry framework},
  author={Sun, Qiming and Berkelbach, Timothy C and Blunt, Nick S and Booth, George H and Guo, Sheng and Li, Zhendong and Liu, Junzi and McClain, James D and Sayfutyarova, Elvira R and Sharma, Sandeep and Wouters, Sebastian and Chan, Garnet Kin-Lic },
  journal={WIREs: Comput. Mol. Sci.},
  volume={8},
  number={1},
  pages={e1340},
  year={2018},
  publisher={Wiley Online Library}
}

@article{bauer2020quantum,
  title={Quantum algorithms for quantum chemistry and quantum materials science},
  author={Bauer, Bela and Bravyi, Sergey and Motta, Mario and Chan, Garnet Kin-Lic},
  journal={Chem. Rev.},
  volume={120},
  number={22},
  pages={12685},
  year={2020},
  publisher={ACS Publications}
}

@article{mcardle2020quantum,
  title={Quantum computational chemistry},
  author={McArdle, Sam and Endo, Suguru and Aspuru-Guzik, Al{\'a}n and Benjamin, Simon C and Yuan, Xiao},
  journal={Rev. Mod. Phys.},
  volume={92},
  number={1},
  pages={015003},
  year={2020},
  publisher={APS}
}

@article{cao2019quantum,
  title={Quantum chemistry in the age of quantum computing},
  author={Cao, Yudong and Romero, Jonathan and Olson, Jonathan P and Degroote, Matthias and Johnson, Peter D and Kieferov{\'a}, M{\'a}ria and Kivlichan, Ian D and Menke, Tim and Peropadre, Borja and Sawaya, Nicolas PD and others},
  journal={Chem. Rev.},
  volume={119},
  number={19},
  pages={10856},
  year={2019},
  publisher={ACS Publications}
}

@article{motta2021emerging,
  title={Emerging quantum computing algorithms for quantum chemistry},
  author={Motta, Mario and Rice, Julia E},
  journal={WIRE: Comput. Mol. Sci.},
  pages={e1580},
  year={2021},
  publisher={Wiley Online Library}
}

@article{de2021materials,
  title={Materials challenges and opportunities for quantum computing hardware},
  author={de Leon, Nathalie P and Itoh, Kohei M and Kim, Dohun and Mehta, Karan K and Northup, Tracy E and Paik, Hanhee and Palmer, BS and Samarth, Nitin and Sangtawesin, Sorawis and Steuerman, DW},
  journal={Science},
  volume={372},
  number={6539},
  pages={eabb2823},
  year={2021},
  publisher={American Association for the Advancement of Science}
}

@article{peruzzo2014variational,
  title={A variational eigenvalue solver on a photonic quantum processor},
  author={Peruzzo, Alberto and McClean, Jarrod and Shadbolt, Peter and Yung, Man-Hong and Zhou, Xiao-Qi and Love, Peter J and Aspuru-Guzik, Al{\'a}n and O’brien, Jeremy L},
  journal={Nat. Commun.},
  volume={5},
  number={1},
  pages={1},
  year={2014},
  publisher={Nature Publishing Group}
}

@article{mcclean2016theory,
  title={The theory of variational hybrid quantum-classical algorithms},
  author={McClean, Jarrod R and Romero, Jonathan and Babbush, Ryan and Aspuru-Guzik, Al{\'a}n},
  journal={New J. Phys.},
  volume={18},
  number={2},
  pages={023023},
  year={2016},
  publisher={IOP Publishing}
}

@article{kandala2017hardware,
  title={Hardware-efficient variational quantum eigensolver for small molecules and quantum magnets},
  author={Kandala, Abhinav and Mezzacapo, Antonio and Temme, Kristan and Takita, Maika and Brink, Markus and Chow, Jerry M and Gambetta, Jay M},
  journal={Nature},
  volume={549},
  number={7671},
  pages={242},
  year={2017},
  publisher={Nature Publishing Group}
}

@article{bartlett1989alternative,
  title={Alternative coupled-cluster ans{\"a}tze II. The unitary coupled-cluster method},
  author={Bartlett, Rodney J and Kucharski, Stanislaw A and Noga, Jozef},
  journal={Chem. Phys. Lett.},
  volume={155},
  number={1},
  pages={133},
  year={1989},
  publisher={Elsevier}
}

@article{taube2006new,
  title={New perspectives on unitary coupled-cluster theory},
  author={Taube, Andrew G and Bartlett, Rodney J},
  journal={Int. J. Quant. Chem.},
  volume={106},
  number={15},
  pages={3393},
  year={2006},
  publisher={Wiley Online Library}
}

@article{romero2018strategies,
  title={Strategies for quantum computing molecular energies using the unitary coupled cluster ansatz},
  author={Romero, Jonathan and Babbush, Ryan and McClean, Jarrod R and Hempel, Cornelius and Love, Peter J and Aspuru-Guzik, Al{\'a}n},
  journal={Quant. Sci. Tech.},
  volume={4},
  number={1},
  pages={014008},
  year={2018},
  publisher={IOP Publishing}
}

@article{anand2022quantum,
  title={A quantum computing view on unitary coupled cluster theory},
  author={Anand, Abhinav and Schleich, Philipp and Alperin-Lea, Sumner and Jensen, Phillip WK and Sim, Sukin and D{\'\i}az-Tinoco, Manuel and Kottmann, Jakob S and Degroote, Matthias and Izmaylov, Artur F and Aspuru-Guzik, Al{\'a}n},
  journal={Chem. Soc. Rev.},
  year={2022},
  volume={51},
  number={5},
  pages={1659},
  publisher={Royal Society of Chemistry}
}

@article{grimsley2019adaptive,
  title={An adaptive variational algorithm for exact molecular simulations on a quantum computer},
  author={Grimsley, Harper R and Economou, Sophia E and Barnes, Edwin and Mayhall, Nicholas J},
  journal={Nat. Commun.},
  volume={10},
  number={1},
  pages={1},
  year={2019},
  publisher={Nature Publishing Group}
}

@article{tang2021qubit,
  title={qubit-adapt-vqe: An adaptive algorithm for constructing hardware-efficient ans{\"a}tze on a quantum processor},
  author={Tang, Ho Lun and Shkolnikov, VO and Barron, George S and Grimsley, Harper R and Mayhall, Nicholas J and Barnes, Edwin and Economou, Sophia E},
  journal={PRX Quantum},
  volume={2},
  number={2},
  pages={020310},
  year={2021},
  publisher={APS}
}

@article{zhang2021mutual,
  title={Mutual information-assisted adaptive variational quantum eigensolver},
  author={Zhang, Zi-Jian and Kyaw, Thi Ha and Kottmann, Jakob S and Degroote, Matthias and Aspuru-Guzik, Al{\'a}n},
  journal={Quant. Sci. Tech.},
  volume={6},
  number={3},
  pages={035001},
  year={2021},
  publisher={IOP Publishing}
}

@article{fac-UCC,
author = {Chen, Jia and Cheng, Hai-Ping and Freericks, J. K.},
title = {Low-Depth Unitary Coupled Cluster Theory for Quantum Computation},
journal = {J. Chem. Theory Comput.},
volume = {18},
number = {4},
pages = {2193},
year = {2022},
}

@article{xu2022decomposition,
  title={Decomposition of high-rank factorized unitary coupled-cluster operators using ancilla and multiqubit controlled low-rank counterparts},
  author={Xu, Luogen and Lee, Joseph T and Freericks, JK},
  journal={Phys. Rev. A},
  volume={105},
  number={1},
  pages={012406},
  year={2022},
  publisher={APS}
}

@article{bauman2019downfolding,
  title={Downfolding of many-body Hamiltonians using active-space models: Extension of the sub-system embedding sub-algebras approach to unitary coupled cluster formalisms},
  author={Bauman, Nicholas P and Bylaska, Eric J and Krishnamoorthy, Sriram and Low, Guang Hao and Wiebe, Nathan and Granade, Christopher E and Roetteler, Martin and Troyer, Matthias and Kowalski, Karol},
  journal={J. Chem. Phys.},
  volume={151},
  number={1},
  pages={014107},
  year={2019},
  publisher={AIP Publishing LLC}
}

@article{kowalski2021dimensionality,
  title={Dimensionality reduction of the many-body problem using coupled-cluster subsystem flow equations: Classical and quantum computing perspective},
  author={Kowalski, Karol},
  journal={Phys. Rev. A},
  volume={104},
  number={3},
  pages={032804},
  year={2021},
  publisher={APS}
}

@article{fujii2022deep,
  title={Deep Variational Quantum Eigensolver: a divide-and-conquer method for solving a larger problem with smaller size quantum computers},
  author={Fujii, Keisuke and Mizuta, Kaoru and Ueda, Hiroshi and Mitarai, Kosuke and Mizukami, Wataru and Nakagawa, Yuya O},
  journal={PRX Quantum},
  volume={3},
  number={1},
  pages={010346},
  year={2022},
  publisher={APS}
}

@article{mizuta2021deep,
  title={Deep variational quantum eigensolver for excited states and its application to quantum chemistry calculation of periodic materials},
  author={Mizuta, Kaoru and Fujii, Mikiya and Fujii, Shigeki and Ichikawa, Kazuhide and Imamura, Yutaka and Okuno, Yukihiro and Nakagawa, Yuya O},
  journal={Phys. Rev. Res.},
  volume={3},
  number={4},
  pages={043121},
  year={2021},
  publisher={APS}
}

@article{mineh2022solving,
  title={Solving the Hubbard model using density matrix embedding theory and the variational quantum eigensolver},
  author={Mineh, Lana and Montanaro, Ashley},
  journal={Phys. Rev. B},
  volume={105},
  number={12},
  pages={125117},
  year={2022},
  publisher={APS}
}

@article{tilly2021reduced,
  title={Reduced density matrix sampling: Self-consistent embedding and multiscale electronic structure on current generation quantum computers},
  author={Tilly, Jules and Sriluckshmy, PV and Patel, Akashkumar and Fontana, Enrico and Rungger, Ivan and Grant, Edward and Anderson, Robert and Tennyson, Jonathan and Booth, George H},
  journal={Phys. Rev. Res.},
  volume={3},
  number={3},
  pages={033230},
  year={2021},
  publisher={APS}
}

@article{keen2020quantum,
  title={Quantum-classical simulation of two-site dynamical mean-field theory on noisy quantum hardware},
  author={Keen, Trevor and Maier, Thomas and Johnston, Steven and Lougovski, Pavel},
  journal={Quant. Sci. Tech.},
  volume={5},
  number={3},
  pages={035001},
  year={2020},
  publisher={IOP Publishing}
}

@article{bauer2016hybrid,
  title={Hybrid quantum-classical approach to correlated materials},
  author={Bauer, Bela and Wecker, Dave and Millis, Andrew J and Hastings, Matthew B and Troyer, Matthias},
  journal={Phys. Rev. X},
  volume={6},
  number={3},
  pages={031045},
  year={2016},
  publisher={APS}
}

@misc{Qiskit,
      author = {MD SAJID ANIS and Abby-Mitchell and H{\'e}ctor Abraham and AduOffei and Rochisha Agarwal and Gabriele Agliardi and Merav Aharoni and Ismail Yunus Akhalwaya and Gadi Aleksandrowicz and Thomas Alexander and Matthew Amy and Sashwat Anagolum and Anthony-Gandon and Eli Arbel and Abraham Asfaw and Anish Athalye and Artur Avkhadiev and Carlos Azaustre and PRATHAMESH BHOLE and Abhik Banerjee and Santanu Banerjee and Will Bang and Aman Bansal and Panagiotis Barkoutsos and Ashish Barnawal and George Barron and George S. Barron and Luciano Bello and Yael Ben-Haim and M. Chandler Bennett and Daniel Bevenius and Dhruv Bhatnagar and Arjun Bhobe and Paolo Bianchini and Lev S. Bishop and Carsten Blank and others},
       title = {Qiskit: An Open-source Framework for Quantum Computing},
       year = {2021},
       doi = {10.5281/zenodo.2573505},
       url = {https://doi.org/10.5281/zenodo.2573505}
}

@article{QCC2018,
  title={Qubit coupled cluster method: a systematic approach to quantum chemistry on a quantum computer},
  author={Ryabinkin, Ilya G and Yen, Tzu-Ching and Genin, Scott N and Izmaylov, Artur F},
  journal={J. Chem. Theory Comput.},
  volume={14},
  number={12},
  pages={6317},
  year={2018},
  publisher={ACS Publications}
}

@article{QCC2020,
  title={Iterative qubit coupled cluster approach with efficient screening of generators},
  author={Ryabinkin, Ilya G and Lang, Robert A and Genin, Scott N and Izmaylov, Artur F},
  journal={J. Chem. Theory Comput.},
  volume={16},
  number={2},
  pages={1055},
  year={2020},
  publisher={ACS Publications}
}

@article{selectiveUCC2022,
  title={Unitary selective coupled-cluster method},
  author={Fedorov, Dmitry A and Alexeev, Yuri and Gray, Stephen K and Otten, Matthew},
  journal={Quantum},
  volume={6},
  pages={703},
  year={2022},
  publisher={Verein zur F{\"o}rderung des Open Access Publizierens in den Quantenwissenschaften}
}

@article{projectiveVQE2021,
  title={Simulating many-body systems with a projective quantum eigensolver},
  author={Stair, Nicholas H and Evangelista, Francesco A},
  journal={PRX Quantum},
  volume={2},
  number={3},
  pages={030301},
  year={2021},
  publisher={APS}
}

@article{bravyi2022future,
  title={The future of quantum computing with superconducting qubits},
  author={Bravyi, Sergei and Dial, Oliver and Gambetta, Jay M and Gil, Dar{\'\i}o 
          and Nazario, Zaira},
  journal={J. Appl. Phys.},
  volume={132},
  pages={160902},
  year={2022},
  publisher={AIP Publishing}
}

@article{kim2023evidence,
  title={Evidence for the utility of quantum computing before fault tolerance},
  author={Kim, Youngseok and Eddins, Andrew and Anand, Sajant and Wei, Ken Xuan and van den Berg, Ewout and Rosenblatt, Sami and Nayfeh, Hasan and Wu, Yantao and Zaletel, Michael and Temme, Kristan and others},
  journal={Nature},
  volume={618},
  pages={500--505},
  year={2023},
  publisher={Nature Publishing Group}
}

@article{Motta2023,
  title = {Bridging physical intuition and hardware efficiency for correlated electronic states: the local unitary cluster Jastrow ansatz for electronic structure},
  author = {Motta,  Mario and Sung,  Kevin J. and Whaley,  K. Birgitta and Head-Gordon,  Martin and Shee,  James},
  journal = {Chemical Science},
  volume = {14},
  number = {40},
  pages = {11213–11227},
  year = {2023},
  publisher = {Royal Society of Chemistry (RSC)}
}

@article{Knizia2013,
  title = {Density Matrix Embedding: A Strong-Coupling Quantum Embedding Theory},
  author = {Knizia,  Gerald and Chan,  Garnet Kin-Lic},
  journal = {J. Chem. Theory Comput.},
  volume = {9},
  number = {3},
  pages = {1428–1432},
  year = {2013},
  publisher = {American Chemical Society (ACS)}
}

@article{Gujarati2023,
  title = {Quantum computation of reactions on surfaces using local embedding},
  author = {Gujarati,  Tanvi P. and Motta,  Mario and Friedhoff,  Triet Nguyen and Rice,  Julia E. and Nguyen,  Nam and Barkoutsos,  Panagiotis Kl. and Thompson,  Richard J. and Smith,  Tyler and Kagele,  Marna and Brei,  Mark and Jones,  Barbara A. and Williams,  Kristen},
  journal = {Npj Quantum Inf},
  volume = {9},
  number = {1},
  year = {2023},
  publisher = {Springer Science and Business Media LLC}
}

@article{Dresselhaus2015,
  title = {Self-consistent embedding of density-matrix renormalization group wavefunctions in a density functional environment},
  author = {Dresselhaus,  Thomas and Neugebauer,  Johannes and Knecht,  Stefan and Keller,  Sebastian and Ma,  Yingjin and Reiher,  Markus},
  journal = {J. Chem. Phys.},
  volume = {142},
  number = {4},
  year = {2015},
  publisher = {AIP Publishing}
}

@article{RobledoMoreno2025,
  title = {Chemistry beyond the scale of exact diagonalization on a quantum-centric supercomputer},
  author = {Robledo-Moreno,  Javier and Motta,  Mario and Haas,  Holger and Javadi-Abhari,  Ali and Jurcevic,  Petar and Kirby,  William and Martiel,  Simon and Sharma,  Kunal and Sharma,  Sandeep and Shirakawa,  Tomonori and Sitdikov,  Iskandar and Sun,  Rong-Yang and Sung,  Kevin J. and Takita,  Maika and Tran,  Minh C. and Yunoki,  Seiji and Mezzacapo,  Antonio},
  journal = {Sci. Adv.},
  volume = {11},
  number = {25},
  year = {2025},
  publisher = {American Association for the Advancement of Science (AAAS)}
}

@article{Barison2025,
  title = {Quantum-centric computation of molecular excited states with extended sample-based quantum diagonalization},
  author = {Barison,  Stefano and Robledo Moreno,  Javier and Motta,  Mario},
  journal = {Quantum Sci. Techno.},
  volume = {10},
  number = {2},
  year = {2025},
  pages = {025034},
  publisher = {IOP Publishing}
}

@article{Danilov2025,
  title = {Enhancing the Accuracy and Efficiency of Sample-Based Quantum Diagonalization with Phaseless Auxiliary-Field Quantum Monte Carlo},
  author = {Danilov,  Don and Robledo-Moreno,  Javier and Sung,  Kevin J. and Motta,  Mario and Shee,  James},
  journal = {J. Chem. Theory Comput.},
  volume = {21},
  number = {22},
  year = {2025},
  pages = {11585–11594},
  publisher = {American Chemical Society (ACS)}
}

@article{Kaliakin2025,
  title = {Implicit Solvent Sample-Based Quantum Diagonalization},
  author = {Kaliakin,  Danil and Shajan,  Akhil and Liang,  Fangchun and Merz,  Kenneth M.},
  journal = {J. Chem. Phys B},
  volume = {129},
  number = {23},
  year = {2025},
  pages = {5788–5796},
  publisher = {American Chemical Society (ACS)}
}

@article{Shajan2025,
  title = {Toward Quantum-Centric Simulations of Extended Molecules: Sample-Based Quantum Diagonalization Enhanced with Density Matrix Embedding Theory},
  author = {Shajan,  Akhil and Kaliakin,  Danil and Mitra,  Abhishek and Robledo Moreno,  Javier and Li,  Zhen and Motta,  Mario and Johnson,  Caleb and Saki,  Abdullah Ash and Das,  Susanta and Sitdikov,  Iskandar and Mezzacapo,  Antonio and Merz,  Kenneth M.},
  journal = {J. Chem. Theory Comput.},
  volume = {21},
  number = {14},
  year = {2025},
  pages = {6801–6810},
  publisher = {American Chemical Society (ACS)}
}

\end{document}